\documentclass[prb,aps,twocolumn,home,floats,showpacs]{revtex4}

\usepackage{graphics}
\usepackage{amsmath}
\usepackage{dcolumn}
\usepackage{citesort}
\begin{document}

\title {{\rm\small\hfill (submitted to Phys. Rev. B)}\\
Density-functional theory study of the initial oxygen incorporation in Pd(111)}

\author{Mira Todorova}
\affiliation{Fritz-Haber-Institut der Max-Planck-Gesellschaft, Faradayweg
4-6, D-14195 Berlin, Germany}

\author{Karsten Reuter}
\affiliation{Fritz-Haber-Institut der Max-Planck-Gesellschaft, Faradayweg
4-6, D-14195 Berlin, Germany}

\author{Matthias Scheffler}
\affiliation{Fritz-Haber-Institut der Max-Planck-Gesellschaft, Faradayweg
4-6, D-14195 Berlin, Germany}

\received{27 December 2004}

\begin{abstract}
Pd(111) has recently been shown to exhibit a propensity to form a sub-nanometer thin surface oxide film already well before a full monolayer coverage of adsorbed O atoms is reached on the surface. Aiming at an atomic-scale understanding of this finding, we study the initial oxygen incorporation into the Pd(111) surface using density-functional theory. We find that oxygen incorporation into the sub-surface region starts at essentially the same coverage as formation of the surface oxide. This implies that the role of 
sub-surface oxygen should be considered as that of a metastable precursor in the oxidation process of the surface. The mechanisms found to play a role towards the ensuing stabilization of an ordered oxidic structure with a mixed 
on-surface/sub-surface site occupation follow a clear trend over the late 
$4d$ transition metal series, as seen by comparing our data to previously 
published studies concerned with oxide formation at the basal surface of Ru, Rh and Ag. The formation of a linearly aligned O-TM-O trilayered structure (TM = Ru, Rh, Pd, Ag), together with an efficient coupling to the underlying substrate seem to be key ingredients in this respect.
\end{abstract}

\pacs{81.65.Mq, 68.43.Bc, 68.47.Gh}


\maketitle

\section{Introduction}

Contact with a rich gas-phase is characteristic for solid surfaces in most applications and real-life situations. The concomitant interaction with the atoms and molecules of the environment can induce significant materials changes, exemplified prominently by the corrosion of metal surfaces. The oxide formation underlying the latter is an everyday occurrence in our oxygen-rich atmosphere and of great importance for the atomic-scale understanding of many technologically relevant processes like oxidation catalysis. There is general agreement that the atomistic steps leading to oxide formation involve dissociative oxygen chemisorption on the metal surfaces, often the formation of  
adsorbate structures, followed by incorporation of oxygen into the surface and finally (or simultaneously) the transformation to oxidic films (surface oxides) and bulk oxide growth. Still, the microscopic understanding of these different processes is quite shallow. The sequence of atomistic events may be quite complex, and rather than successively, they may occur simultaneously or even concertedly, depending crucially on temperature and partial pressures in the surrounding.

Investigations on well-defined single-crystals help to improve this atomic-scale insight, and due to its widespread technological use, Pd is an appropriate target material. The initial chemisorption of oxygen on Pd single crystal surfaces has correspondingly been the subject of intensive experimental research, e.g. Refs. \onlinecite{conrad77, legare81, imbihl86, guo89, banse90, voogt97, seitsonen00, lundgren02}. The picture arising from these studies for the close packed (111) surface is that for temperatures $T > 200$\,K oxygen adsorption on the surface is dissociative. An ordered $(2 \times 2)$ overlayer is observed at a coverage of $\Theta = 0.25$ monolayers (ML) \cite{conrad77}, while at higher coverages on-surface adsorption seems to compete with the formation of oxidic structures: depending on the preparation conditions a $({\sqrt 3} \times {\sqrt 3})\,R30^{\rm o}$ \cite{conrad77}, a $(1 \times 1)$ \cite{legare81, voogt97} and a ``complex'' structure \cite{conrad77, lundgren02} have been observed, where the ``complex'' structure with $\Theta = 0.7$\,ML was recently characterized as a pseudo-commensurate surface oxide consisting of an O-Pd-O trilayered geometry on the Pd(111) substrate \cite{lundgren02}.

In a previous publication \cite{todorova04} we discussed the on-surface adsorption of oxygen on the Pd(111) surface for coverages up to one monolayer.  Using density-functional theory (DFT) we identified the fcc hollow sites as the most stable adsorption sites \cite{seitsonen00}, which is consistent with a  trend observed for the basal surface of the other late $4d$ transition metals, namely Ru, Rh and Ag, for which oxygen also adsorbs preferentially in those sites following the bulk-stacking sequence. Furthermore, we found that overall repulsive lateral interactions in the addressed coverage range give rise to a pronounced reduction in the adsorption energy with increasing oxygen coverage, but are still clearly exothermic up to a full ML $(1 \times 1)$ phase. 
In the present paper we now investigate the initial oxygen incorporation into 
the palladium (111) surface using DFT. We find that occupation of sub-surface 
sites becomes energetically more favorable than continued adsorption into 
on-surface sites already at total coverages well below 1\,ML. We analyze the ensuing structures and discuss what might be the mechanisms which make one structure more stable than another. A comparison to the $4d$ neighbors of Pd in the periodic table shows that the mechanisms found here are valid there, as well. Last, but not least, we address the question of how much oxygen can be dissolved into a palladium crystal, an issue which has often been the object of speculation in the experimental literature.

\section{Theory}\label{theory}

The DFT calculations are performed within the full-potential linear augmented plane wave (FP-LAPW) method \cite{blaha99,kohler96,petersen00} using the generalized gradient approximation (GGA) \cite{perdew96} for the 
exchange-correlation functional. All surface structures are modeled in a supercell geometry, employing a symmetric slab consisting of seven (111) Pd layers. A vacuum region of 17{\AA} ensures the decoupling of consecutive slabs. All atomic positions within the adsorbed oxygen layers and the two outermost Pd layers are fully relaxed.

The FP-LAPW basis-set parameters are as follows: the muffin-tin radii are $R_{\rm{MT}}^{\rm{Pd}} = 2.25$\,bohr and $R_{\rm{MT}}^{\rm{O}} = 1.3$\,bohr. A wave function expansion inside the muffin tin spheres up to $l_{\rm{max}}^{\rm{wf}} = 12$ and a potential expansion up to $l_{\rm{max}}^{\rm{pot}} = 4$ is used. The $4s$ and $4p$ semicore states of Pd, as well as the O $2s$ states are described by adding local orbitals to the 
FP-LAPW basis set. The energy cutoff for the plane wave representation in the interstitial region between the muffin tin spheres is $E^{\rm max}_{\rm wf} = 17$\,Ry for the wavefunctions and $E^{\rm max}_{\rm pot} = 169$\,Ry for the potential. A $(12 \times 12 \times 1)$ Monkhorst-Pack grid with $19$ ${\bf k}$ points in the irreducible wedge is employed for the Brillouin zone integration in a $(1 \times 1)$ cell. To obtain the same sampling of the reciprocal space for bigger surface cells, this number is reduced accordingly.

This is the same basis set as in our preceding publication on the oxygen 
on-surface adsorption on Pd(111) \cite{todorova04}. Its high accuracy has already been described there, and with respect to the present work it is primarily characterized by a $\pm 30$\,meV ($\pm 50$\,meV) numerical uncertainty, when comparing relative binding energies in geometries containing an equal (unequal) amount of O atoms.

\section{Results}

\subsection{Initial oxygen incorporation}

This subsection addresses the initial oxygen incorporation into the surface by studying the stability and properties of a set of structures with oxygen coverages in the range $0 < \Theta_{\rm tot} \leq 1$ML.  We employ $(2 \times 2)$ unit cells and calculate the average binding energy per O atom with respect to the dissociation energy of an oxygen molecule. Our sign convention is such that positive binding energies indicate that adsorption is exothermic. From one ($\Theta_{\rm tot} = 0.25$\,ML) up to four ($\Theta_{\rm tot} = 1.00$\,ML) oxygen atoms are placed in any combinatorial way that is possible in a $(2 \times 2)$ unit cell in on-surface sites and in interstitial sites between the first and second substrate layers. Occupation of sub-surface sites will then obviously commence when a structure containing oxygen in sub-surface sites becomes energetically more stable than its counterpart with the same total adsorbate coverage, $\Theta_{\rm tot}$, but all oxygen atoms located on the surface.

\begin{figure*}[t!]
\scalebox{0.6}{\includegraphics{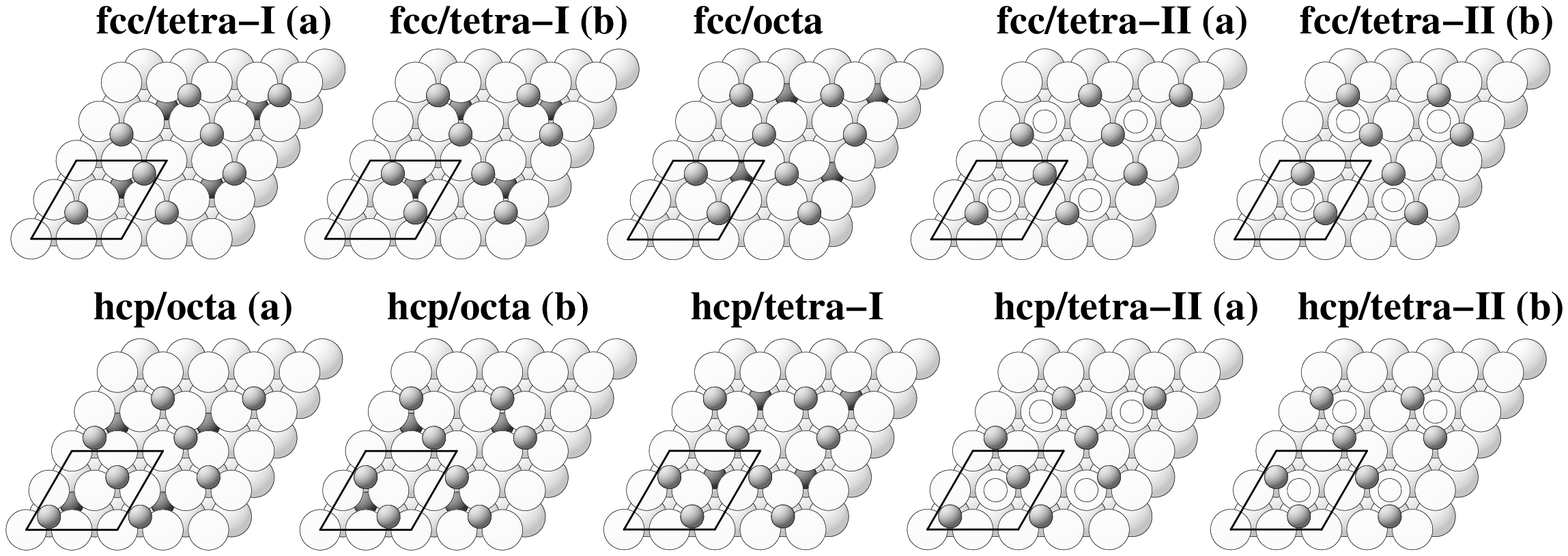}}
\caption{Top view of all possible on-surface/sub-surface site combinations for a total coverage $\Theta_{\rm tot}$ = 0.75ML in a $(2 \times 2)$ arrangement, and one of the three oxygen atoms below the surface. Inequivalent geometries with the same site occupation are denoted with (a) and (b), the (a)-geometries being the more stable ones. Big spheres correspond to palladium (white = surface layer, light grey = second layer), small spheres to oxygen (gray = on-surface, black = sub-surface). Sub-surface oxygen atoms in tetra-II sites are invisible in this plot and are indicated schematically by small white circles.}
\label{fig1}
\end{figure*}

In our previous publication on oxygen on Pd(111) \cite{todorova04} we showed that the two threefold hollow sites, that are present on this basal surface, are energetically by far the most stable adsorption sites. Specifically, these are the fcc site, located above a palladium atom of the third layer, and the hcp site, directly above an atom of the second layer. Concerning O incorporation into the Pd(111) surface, we will focus initially on the high-symmetry interstitial sites directly below the topmost Pd(1111) layer, postponing the discussion of deeper sites to the end of this paper. There are three such interstitial sites, namely an octahedral (octa) and a tetrahedral (tetra-I) site directly below the fcc and hcp on-surface sites, respectively, as well as a second tetrahedral (tetra-II) site located below a Pd atom of the first substrate layer. The octa site is coordinated to six palladium atoms, while each of the tetrahedral sites is fourfold coordinated. The distinction between the two tetrahedral sites is due to the presence of the surface; in the bulk, they would be equivalent.

\begin{figure}
\scalebox{0.39}{\includegraphics{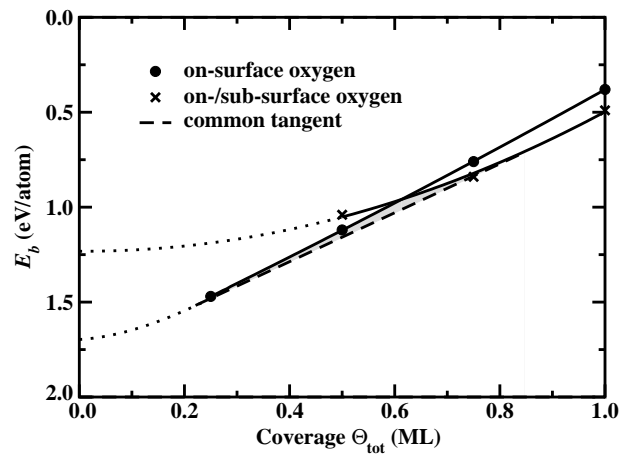}}
\caption{Average binding energy $E_b$ as a function of the total oxygen coverage, with on-surface oxygen in fcc and sub-surface oxygen in tetra-I site. 
The highest $E_b$ is found for the pure on-surface chemisorption phase at
$\Theta_{\rm tot}= 0.25$\,ML, but for $\Theta_{\rm tot}> 0.50$\,ML the 
on-/sub-surface structures become more stable. A coexistence of the on-surface and on-/sub-surface structures is expected in the grey shaded region (s. text). The marks denote the actually calculated binding energies listed in Table \ref{tab1}. The lines are drawn to guide the eye and are schematically extrapolated to zero. The dashed line is the Gibbs construction, i.e. the common tangent.}
\label{fig2}
\end{figure}

The existence of three sub-surface sites together with two energetically favorable threefold hollow sites for on-surface adsorption accounts for a vast host of possible structural combinations that have to be considered. This is exemplified in Fig. \ref{fig1}, which shows all on-surface/sub-surface site combinations resulting for a total coverage of $\Theta_{\rm tot} = 0.75$\,ML within our set of $(2 \times 2)$ geometries, and one of the three O atoms located below the surface. Corresponding structures (not shown) for a total coverage of $\Theta_{\rm tot} = 0.50$\,ML (2 O atoms, 1 of which below the surface) and of $\Theta_{\rm tot} = 1.00$\,ML (4 O atoms, 1 of which below the surface) were tested as well. Very often two symmetry inequivalent geometries exist for the same on-surface/sub-surface coverage and site occupation, as is also illustrated in Fig. \ref{fig1}. We considered both cases, but found that in some situations one of them is energetically very unfavorable: For both the fcc/octa and hcp/tetra-I site combinations one of the possible structures contains a sub-surface oxygen atom located directly below an on-surface oxygen atom at a close distance. Due to the ensuing strong repulsion between the electronegative oxygens (cf. Ref. \onlinecite{todorova04}), these geometries always turn out significantly less stable than all other possibilities. Electrostatic repulsion between the oxygen atoms seems also to play a significant role for the energetic preference of all other structures. For each pair of symmetry inequivalent site combinations, the structure in which the oxygen atoms are located further away from each other (within the constraints of a $(2 \times 2)$ arrangement, cf. Fig. \ref{fig2}) is found to be the more stable one. The coordination offered by the sub-surface site is another decisive factor, since in most cases the tetrahedral sites are favored over the octahedral one.

\begin{table}
\caption{\label{tab1}
Binding energies, $E_b$, (in eV/atom) for the on-surface adsorption in fcc sites and pure sub-surface adsorption into the three interstitial sites as a function of coverage. Additionally shown are the binding energies of the mixed 
on-surface/sub-surface occupation structure with O in fcc/tetra-I sites. Positive binding energies indicate that adsorption is exothermic.} 
\begin{ruledtabular}
\begin{tabular}{c|ccccccc}
Sites           & \multicolumn{7}{c}{$\Theta_{\rm tot}$ (ML)}\\ 
on-/sub-surface &  0.25 & &  0.50 & &  0.75 & &  1.00 \\ \hline
fcc/-           &  1.47 & &  1.12 & &  0.76 & &  0.38 \\ \hline 
 -/tetra-I      & -0.37 & & -0.22 & & -0.13 & & -0.04 \\ 
 -/tetra-II     & -0.58 & & -0.34 & & -0.05 & &  0.10 \\
 -/octa         & -0.78 & & -0.52 & & -0.34 & & -0.38 \\ \hline
fcc/tetra-I     &   -   & &  1.04 & &  0.84 & &  0.49 \\  
\end{tabular}
\end{ruledtabular}
\end{table}

Within the set of structures with mixed on-surface/sub-surface site occupation  
the fcc/tetra-I geometry resulted then as the most stable one at all considered 
coverages ($\Theta_{\rm tot} = 0.50$\,ML, $0.75$\,ML or $1.00$\,ML). While initially pure on-surface adsorption is preferred, already at a total coverage of $\Theta_{\rm tot} = 0.75$\,ML the binding energy of the mixed $(2 \times 2)-(2\,{\rm O_{\rm fcc}} + {\rm O_{\rm tetra-I}})$/Pd(111) structure exceeds the one of the $(2 \times 2)-3\,{\rm O_{\rm fcc}}$/Pd(111) geometry containing the same number of oxygen atoms. Since fcc is the most stable adsorption site for on-surface adsorption \cite{todorova04}, this means that at the latest at $0.75$\,ML total oxygen coverage, structures with oxygen incorporated into the sub-surface region become the favored ones. Figure \ref{fig2} gives a graphical summary of these results (cf. Table \ref{tab1}), where the dashed line marks additionally a Gibbs construction, i.e. the common tangent. In the coverage range spanned by the resulting shaded region (where the common tangent yields the lowest energy) we expect coexistence of the (fcc/-) and the (fcc/tetra-I) structures. This coexistence may result in different domains of these two phases, or it may result in a mixed phase built from the elements of both structures. Obviously the coverage range of coexistence will be enlarged with temperature, and as said above, surface oxides will start to form as well.
A quantitative evaluation of this phase transition and the corresponding actual composition of the surface requires an analysis using, e.g., Monte Carlo simulations (see e.g. Refs. \onlinecite{reuter05, borg05}), which is, however, outside the scope of the present paper.

Apart from coexistence, the determination of the threshold coverage $\Theta_c$ for oxygen penetration from the results summarized in Table \ref{tab1} and Fig. \ref{fig2} has also to be seen in light of the limited set of $(2 \times 2)$ 
surface unit cell structures considered. A quantitative determination of
$\Theta_c$ would require a much larger set of structures (periodic and disordered). For the present study, we therefore only conclude that $\Theta_c$ is roughly at 0.5\,ML total coverage. Interestingly, the coverage above which bulk oxide formation becomes thermodynamically stable is also $\sim$\,1/2\,ML,\cite{todorova02} and surface oxide formation will even set in earlier \cite{reuter04b}. We believe that this points at an important bottleneck function of the initial O incorporation in the oxidation sequence of this surface: Sub-surface O can apparently not exist as a stable phase in the coverage range below which oxide formation would already set in, which in turn implies that the role of sub-surface oxygen is that of a metastable precursor. At not too low temperatures the phase transition to oxidic structures will then proceed nearly instantaneously after the first incorporation of oxygen below the surface.

\begin{figure}
\scalebox{0.40}{\includegraphics{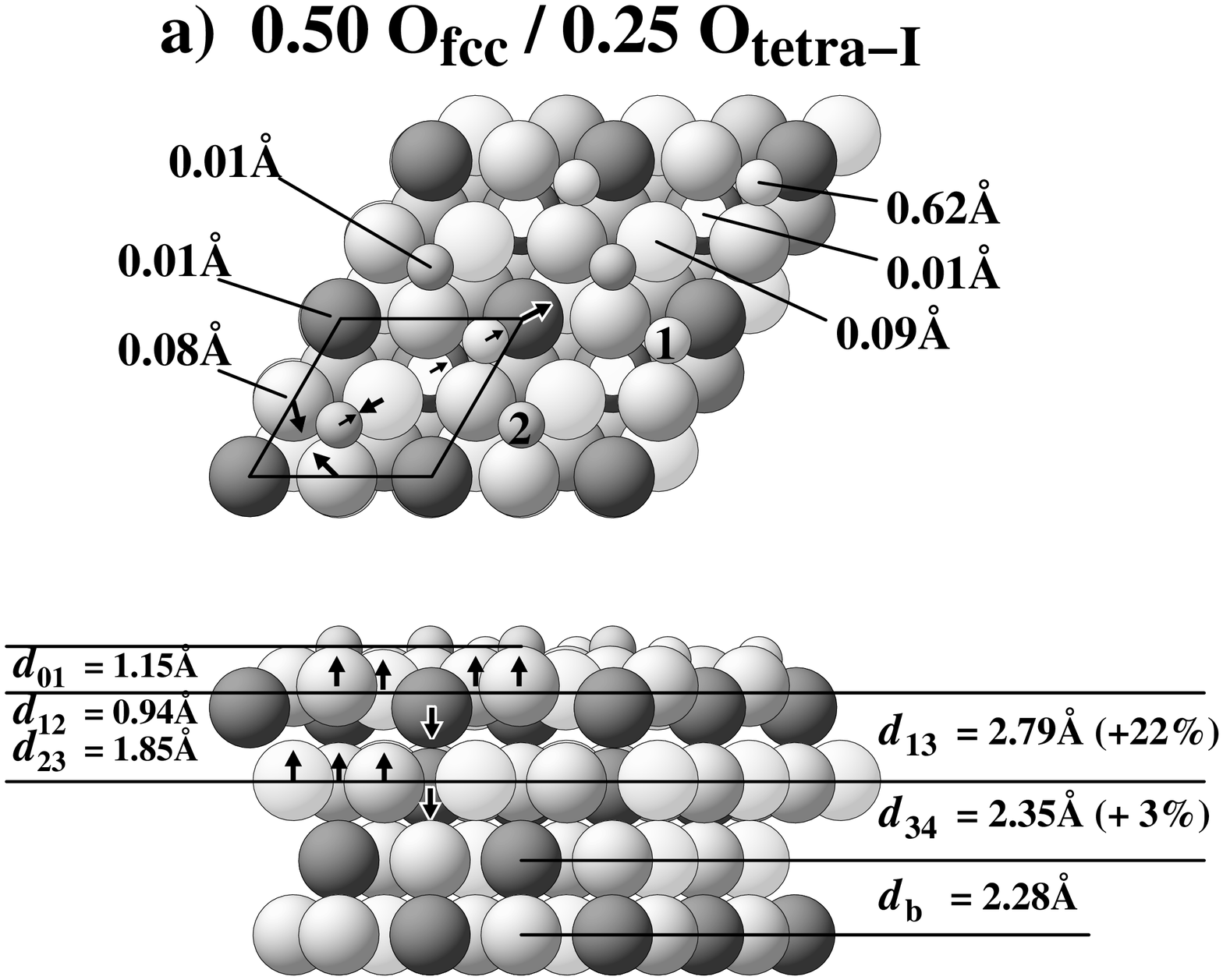}}
\scalebox{0.40}{\includegraphics{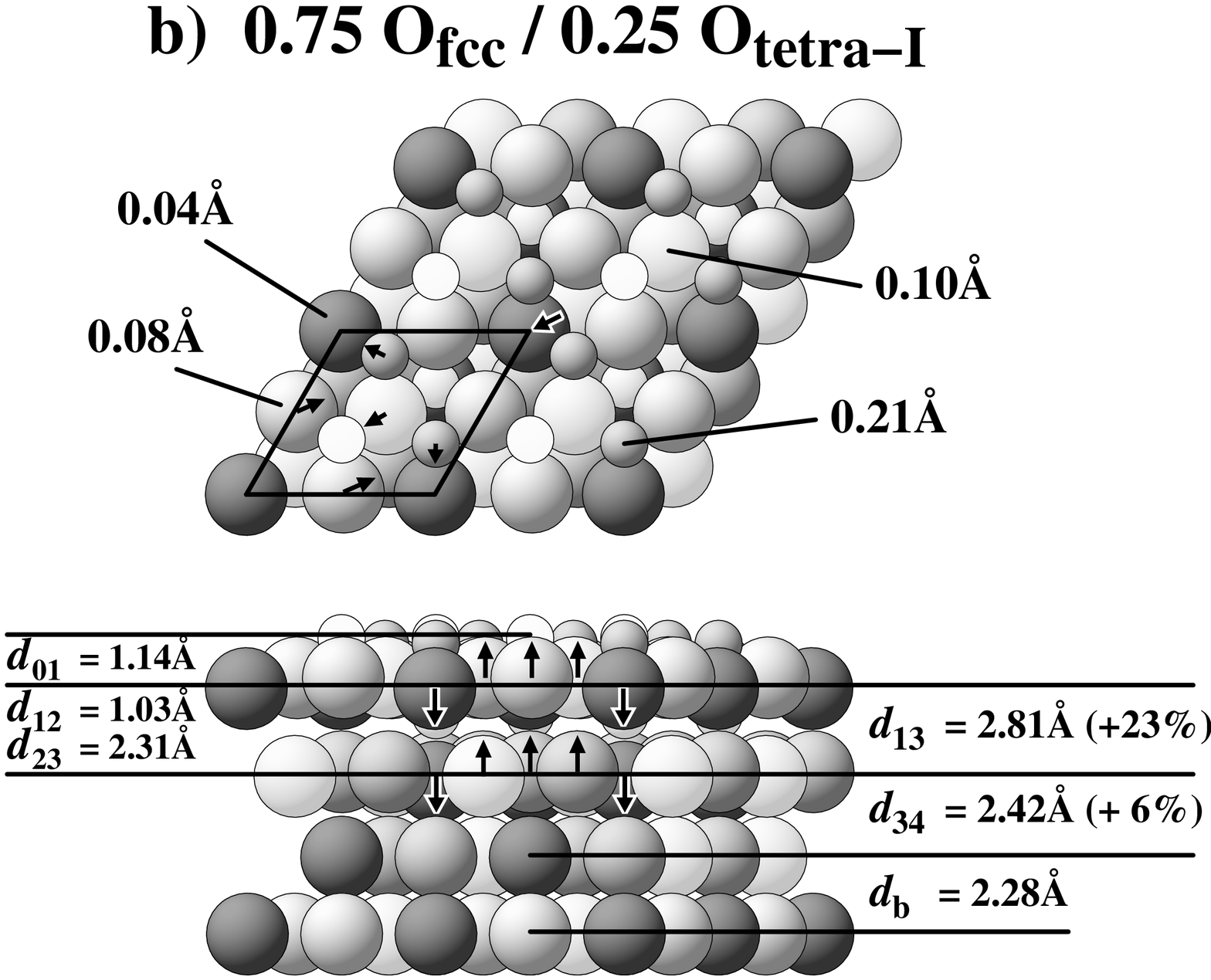}}
\caption{Top view and side view of the most stable structure at (a) 
$\Theta_{\rm tot} = 0.75$\,ML, i.e $(2 \times 2)-(2\,{\rm O}_{\rm fcc}+ {\rm O}_{\rm tetra-I})$ and (b) $\Theta_{\rm tot} = 1.00$\,ML, i.e. $(2 \times 2)-(3\,{\rm O}_{\rm fcc}+ {\rm O}_{\rm tetra-I})$. The small spheres represent the oxygen atoms, while the palladiums are depicted by the large spheres. Atoms, which are lying in the same plane and are equivalent under the present mirror symmetry have the same color. The arrows (not drawn to scale) indicate the direction of the displacement of the substrate atoms from their ideal 
(bulk-like) positions.} 
\label{fig3}
\end{figure}

A rationalization for this low stability of sub-surface oxygen can be found by analyzing the geometric structure of configurations containing sub-surface O in more detail. The atomic geometry of the two most stable structures for $\Theta_{\rm tot} = 0.75$\,ML and $\Theta_{\rm tot} = 1.00$\,ML is displayed in Figs. \ref{fig3}a and \ref{fig3}b. Due to the presence of sub-surface oxygen lowering the symmetry, these geometries exhibit a much more complex relaxation pattern compared to structures with pure on-surface chemisorption, cf. Ref. \onlinecite{todorova04}. However, a number of similarities in the relaxation pattern of different atoms in the two structures can be discerned. The three first-layer palladium atoms in the vicinity of the sub-surface oxygen move sideward away from it and are lifted up, increasing the volume which the 
sub-surface O atom has at its disposal. The fourth palladium atom is pushed away and downward, as the on-surface oxygen atoms in the immediate neighborhood of the sub-surface oxygen try to increase their distance to the latter, moving towards an on-top position. The slight change in the position of the other
on-surface oxygen atom seems then rather as a consequence of this displacement  of the surrounding palladium atoms.

The vertical relaxations shown in Fig. \ref{fig3} for both the $(2 \times 2)-(2\,{\rm O}_{\rm fcc}+ {\rm O}_{\rm tetra-I})$/Pd(111) and the $(2 \times 2)-(3\,{\rm O}_{\rm fcc}+ {\rm O}_{\rm tetra-I})$/Pd(111) system are calculated with respect to the center of mass of each layer. In addition, there is a buckling of $\Delta z_1$ = 0.28\,{\AA} for the first and $\Delta z_2$ = 0.36\,{\AA} for the second layer, respectively, and identical for both geometries. The most prominent feature is, however, a pronounced first layer expansion, which is slightly smaller for the lower coverage case, i.e  $(2 \times 2)-(2\,{\rm O}_{\rm fcc} + {\rm O}_{\rm tetra-I})$/Pd(111), reflecting the reduced on-surface oxygen coverage. Interestingly, the bondlength between the sub-surface O atom and its Pd neighbors remains almost constant for both relaxed structures and takes values between 1.9\,{\AA} and 2\,{\AA}. The corresponding numbers found for on-surface O atoms in our preceding study \cite{todorova04} were in the same range. In this light, we interpret the described relaxation patterns of the mixed on-surface/sub-surface structures largely as a consequence of a local expansion of the crystal lattice to accommodate the sub-surface oxygen at an optimal O-Pd bondlength. 

Comparing these results for the initial penetration of oxygen into the Pd(111) 
surface with corresponding previous DFT studies at the basal surface of the left 
and right neighbors of Pd in the periodic table \cite{reuter02, reuter02a, 
pirovano02, li03}, we are able to identify some clear trends \cite{todorova02}. Concerning the critical coverage at which incorporation commences, occupation of sub-surface sites starts at Ru(0001) and Rh(111) approximately at a full monolayer adsorbed at the surface, while at Ag(111) this happens already at about just a quarter of a monolayer. We have shown \cite {todorova02} that this clear trend is consistent with the bulk properties of these late $4d$ transition metal elements, i.e it scales roughly with their bulk cohesive energies or bulk  
moduli. The already mentioned association to coverages at which oxide formation has been observed also experimentally \cite{lundgren02, boettcher97, over00, 
gustafson04, carlisle00b}, further suggests that sub-surface oxygen acts as a metastable precursor to the formation of oxides at all of these metals. The  sub-surface oxygen incorporation goes in all cases hand in hand with an appreciable deformation of the substrate lattice. Incorporation of 1\,ML oxygen in a tetra-I site causes e.g. relaxations of +53\,\% on Ru(0001), +48\,\% on Rh(111), +46\,\% on Pd(111) and +42\,\% on Ag(111) \cite{reuter02, reuter02a, pirovano02, li03}. Such a distortion is very costly for a comparably hard material like Ru, while it becomes less so for the more noble metals towards the right of the transition metal series, i.e. towards silver. This is also consistent with the pronounced tendency towards sub-surface island formation computed for Ru(0001) \cite{reuter02a} and Rh(111) \cite{pirovano02}, which we find to be much weaker for Pd and Ag \cite{li03}. The stability of oxygen in a sub-surface interstitial site is therefore mainly characterized by minimizing the repulsive interactions with other nearby O atoms, in competition with a reduction of the energy cost for the substrate lattice deformation achieved by locating the O atoms close to each other. Sharing the high cost for the lattice expansion leads in Ru(0001) and Rh(111) to clustering of the oxygen atoms in the sub-surface region. In Pd(111) and Ag(111) the smaller cost reduces this tendency, but leads still to a lower stability of sub-surface O compared to 
on-surface adsorption.

\subsection{Electronic structure}

\begin{figure}
\scalebox{0.31}{\includegraphics{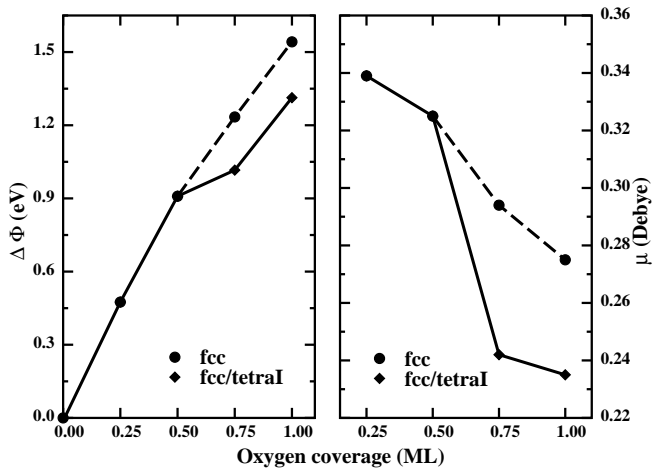}}
\caption{Dependence of the calculated work function (left) and the dipole moment (right) on the total oxygen coverage for the most stable geometry at each considered coverage. At $\Theta_{\rm tot} = 0.75$\,ML and $\Theta_{\rm tot} = 1.00$\,ML both the fcc and the fcc/tetra-I structures are shown. Neither the coexistence phase, nor surface oxides are considered here.}
\label{fig5}
\end{figure}

Turning towards the electronic structure, we first investigate the coverage dependence of the work function and compare the cases of pure on-surface chemisorption to the mixed on-/sub-surface site occupation geometries.  The change in work function upon increasing oxygen coverage relative to the clean surface ($\Phi^{\rm clean} = 5.25$\,eV) is shown in Fig. \ref{fig5}. As expected for an electronegative adsorbate, accomodating oxygen on the surface strongly increases the work function, while the incorporation of oxygen into the 
sub-surface region clearly lowers the work function compared to the pure 
on-surface adsorption structures. This difference remains almost constant, $\approx$ 0.2\,eV for all mixed fcc/tetra-I structures, and reflects the  competition arising between sub-surface and on-surface oxygen atoms for the  bonding charge of the surface Pd atoms. This renders the on-surface O species slightly less negatively charged (as can be seen by comparing the initial state shift of the O\,$1s$ core levels), with respect to a situation with no 
sub-surface oxygen involved. The concomitant depolarization can be quantified via the dipole moment $\mu$, which is related to the adsorbate-induced work function change by the Helmholtz equation
\begin{equation}
\mu = (1/12 \pi) A \Delta \Phi/\Theta\,.
\end{equation}
Here the surface dipole moment is measured in Debye, $A$ is the area per $(1 \times 1)$ surface unit cell (in {\AA}$^2$), the induced work function change, $\Delta\Phi(\Theta) = \Phi(\Theta) - \Phi^{\rm clean}$, is given in eV, and the coverage $\Theta$ in ML. As apparent from Fig. \ref{fig5}, a strong and inward pointing dipole moment is obtained for all considered coverages, consistent with the strong polarization of the density at the surface oxygen atom(s). Concerning the pure on-surface adsorption the observed reduction of the dipole moment with coverage is then a consequence of the growing repulsive interactions within a more and more densely packed overlayer, which induces a slight depolarization of the O atoms \cite{todorova04}. The decrease of $\mu$ for the mixed 
on-surface/sub-surface geometries, can be understood along similar lines. Now, the polarization goes in the other direction and yields a dipole moment contribution of the opposite sign, reducing the overall $\mu$ of the surface. 

\begin{figure}
\scalebox{0.46}{\includegraphics{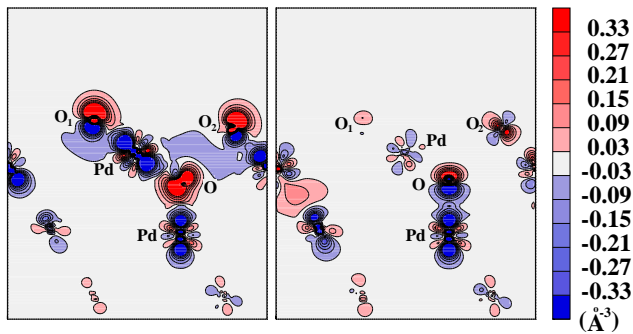}}
\caption{Plots of the {\em difference electron density} for the 
$(2 \times 2) - (2{\rm O}_{\rm fcc} + {\rm O}_{\rm tetra-I})$/Pd(111) structure visualizing the bonding within (left) and the coupling of the O-Pd-O trilayer to the Pd(111) substrate (right). The contour plot depicts the $[{\bar 2}11]$ plane perpendicular to the (111) surface. Regions of electron accumulation and depletion have positive and negative signs, respectively. Contour lines are drawn at 0.06 {\AA}$^{-3}$ spacings. The nomenclature O$_1$ and O$_2$ refers to the two on-surface oxygen atoms marked as ``1'' and ``2'' in Fig. \ref{fig3}.}
\label{fig6}
\end{figure}

To gain further insight into the bonding mechanism underlying the superior stability of the ${\rm O_{fcc}/O_{tetra-I}}$ structures, it is helpful to also analyze the {\em difference electron density}. This quantity is obtained by subtracting from the electron density of the O/Pd(111) system both the electron density of the clean Pd(111) surface and that of the isolated oxygen layers, each time keeping the atoms fixed at the same positions. Figure \ref{fig6} (left) exemplifies this for the $(2\times 2)-(2\,{\rm O_{fcc} + 
O_{tetra-I}})$/Pd(111) geometry. A strong depletion of the $4d_{xz,yz}$ states of the surface palladium atom(s) bonded to both ${\rm O_{fcc}}$ and ${\rm O_{tetra-I}}$ is discernible, while the electron density at the oxygen atoms is polarized. Comparing with equivalent plots for other on-surface/sub-surface structures, it appears that the linear ${\rm O_{fcc} - Pd - O_{tetra-I}}$ coordination furthers the strong hybridization of the Pd\,$4d_{xz,yz}$ orbital with the ${\rm O_{fcc}}$ {\em and} the ${\rm O_{tetra-I}}$ species. Simultaneously, the Pd atom in between effectively screens the electrostatic interactions between the two O atoms, resulting overall in the high stability of this structural arrangement.

Interestingly, a similar linear arrangement O-TM-O (TM = Ru, Rh, Ag) was also observed in the structures with mixed on-/sub-surface site occupation that were reported as most favorable at the basal surface of the neighboring elements of Pd in the periodic system of elements \cite{reuter02a, pirovano02, li03}. For both the fcc-structured Rh and Ag, the most stable mixed on-/sub-surface 
site occupation geometry also consists of a fcc/tetra-I site combination 
\cite{pirovano02, li03}. For ruthenium \cite{reuter02a}, a hcp material, initially a similarly linear hcp/octa structure seems to be the most stable one. A barrierless displacement along the $[{\bar 1} {\bar 1} 2]$ direction leads to an energy gain of about $0.2$\,eV and results then also in a fcc/tetra-I arrangement of the oxygen atoms, with the first Ru layer lying in a stacking fault position. This comparison indicates that the mechanisms playing a role towards stabilizing one mixed site structure over another are quite similar for all these late $4d$ transition metals.

It has been suggested that an improved coupling to the underlying substrate is ultimately the reason for the preference of the fcc/tetra-I combination over the hcp/octa one \cite{reuter02a}. The {\it difference electron density} plot for palladium (Fig. \ref{fig6}) also indicates the presence of such a strong coupling between the O-Pd-O  ``trilayer'' and the underlying palladium atoms. However, this coupling is even better visualized by another type of {\it difference electron density}. Here the electron density of the trilayer and the underlying Pd(111) substrate are subtracted from the overall density of the full structure, as exemplified for the $(2\times 2)-(2\,{\rm O_{fcc} + O_{tetra-I}})$/Pd(111) structure in Fig. \ref{fig6} (right). The apparent coupling, which seems to be mediated predominantly by the hybridization of the Pd\,$4d_{z^2}$ orbital with the O\,$2s$ orbitals of the sub-surface oxygen atom, can be also quantified by 
\begin{equation}
E_{\rm coupling} =\frac{1}{2}(E^{\rm all} - E^{\rm O-Pd-O} - E^{\rm sub})\; .
\label{coupling}
\end{equation}
Here $E^{\rm all}$ is the energy of the whole system, $E^{\rm O-Pd-O}$ is the energy of the trilayer and $E^{\rm sub}$ is the energy of the underlying Pd(111) substrate, as described above. The division by two accounts for the two surfaces present in our symmetric slab geometry. With this we determine the coupling of the trilayer to the substrate as 127\,meV/{\AA}$^2$ in the $(2 \times 2)-(2\,{\rm O}_{\rm fcc}+ {\rm O}_{\rm tetra-I})$/Pd(111) structure and as 142\,meV/{\AA}$^2$ in the $(2 \times 2)- (3\,{\rm O}_{\rm fcc}+ {\rm O}_{\rm tetra-I})$/Pd(111) structure. In the other on-surface/sub-surface structural combinations, the coupling is much weaker, especially for the alternative linear arrangement with hcp/octa site occupations. In the $(2\times 2)-(3\,{\rm O_{hcp} + O_{octa}})$/Pd(111) the coupling is e.g. only 56\,meV/{\AA}$^2$, i.e. less than half of the value in the fcc/tetra-I structure at the same coverage. After maximum distance between O atoms and a linear on-surface/sub-surface arrangement, this confirms the coupling to the underlying substrate as another influential quantity determining the stability of these structures. The ``coupling difference density plot'' in Fig. \ref{fig6} (right) reveals furthermore that most of this coupling is the result of the interactions between the sub-surface O atoms and the second layer Pd atoms, while the first layer Pd atoms (i.e. the ones within the trilayer) contribute only little. Coupling to the underlying substrate has also been identified as playing an important role in the stabilization of the surface oxide structures on both the Pd(111) \cite{lundgren02} and the Pd(100) \cite{todorova03} surfaces. For these two, as well as for the surface oxide identified on the Ag(111) surface \cite{michaelides03,li03b}, oxygen atoms are similarly present at the interface to the underlying substrate. In the case of Pd(100) the coupling has been calculated as 100\,meV/{\AA}$^2$, \cite{todorova03} i.e. of the same order of magnitude as now found for the most stable fcc/tetra-I geometries.

\subsection{Bulk dissolved oxygen}

So far we have investigated the presence of sub-surface oxygen between the first and the second layer of the Pd(111) surface. It is conceivable that oxygen can also be incorporated between deeper lying layers, i.e layers which are further away from the surface. The incorporation of oxygen into the bulk material, on which we will focus in the following, is the limiting case for such a penetration, and will furthermore allow us to estimate the amount of dissolved O atoms in a palladium crystal. To address this question the binding energy of an oxygen atom in an octahedral interstitial site has been calculated. Considering the much smaller volume in the tetrahedral interstitial sites and the above described strong lateral expansion of the crystal lattice to achieve optimum sub-surface O-Pd bond lengths of the order of 2\,{\AA}, we expect the tetrahedral sites to be significantly less stable (and thus less relevant for the total uptake) than the octahedral ones. Concentrating therefore on the octahedral sites, we employed a large $(4 \times 4 \times 6)$ palladium bulk unit cell with 97 atoms and allowed relaxation of the nearest Pd neighbor shell around the interstitial site. Due to the still finite size of the supercell long range elastic interactions of the metallic lattice are not properly accounted for in the resulting binding energy of $E_b = -0.70$\,eV. In a similar study concerning the oxygen concentration in a ruthenium crystal \cite{reuter02a} the long range elastic interactions were estimated to increase the calculated binding energy by no more than $0.5$\,eV. The calculated bulk modulus of ruthenium ($B_0 = 320$\,GPa) is nearly twice as large as that of palladium ($B_0 = 163$\,GPa). This leads us to assume that a consideration of the long range elastic interactions will probably not improve the calculated binding energy by more than $0.30$\,eV, leading to a conservative upper limit for the binding energy of bulk dissolved oxygen of $E_b \approx -0.4$\,eV. With respect to gas phase O$_2$, occupation of interstitial sites is therefore endothermic. Still, a finite concentration can nevertheless be expected at finite temperatures purely on entropic grounds. Together with the vast number of available interstitial sites, a considerable amount of oxygen might therefore still be deposited into a macroscopic sample, which we estimate with the following thermodynamic reasoning:

In thermodynamic equilibrium the number of oxygen atoms, $N_{\rm O}$, that are dissolved in the bulk, depends on the total number of available interstitial sites, $N$, in the sample, as well as on the temperature $T$ and pressure $p_{\rm O_2}$ in the surrounding. By minimizing the Gibbs free energy one 
obtains for the concentration of oxygen in the Pd bulk \cite{ashcroft76}
\begin{equation}
\frac{N_{\rm O}}{N} = e^{[-E_b+\Delta\mu_{\rm O}(T,p_{\rm O_2})]/k_{\rm B}T},
\label{eq1}
\end{equation}
where $k_{\rm B}$ is the Boltzmann constant, $E_b$ the binding energy with 
respect to half of the total energy of molecular oxygen (positive for exothermicity) and the gas phase conditions are summarized in the chemical potential, $\Delta\mu_{\rm O}(T,p_{\rm O_2})$, which is also referred to $\frac{1}{2} {\rm O}_2$. The latter quantity is related to gas phase temperature and pressure via ideal gas laws \cite{reuter02b}
\begin{equation}
\begin{split}
\Delta \mu_{\rm O} (T,p_{\rm O_2})= \frac{1}{2} \biggl[ \tilde{\mu}_{{\rm O}_2}(T,p^0) + k_B T ln \biggl(\frac{p_{{\rm O}_2}}{p^0}\biggl)\biggl],
\end{split}
\end{equation}
where the temperature dependence of $\tilde{\mu}_{{\rm O}_2}(T,p^0)$ includes 
the contributions from rotations and vibrations of the molecules, as well as the 
ideal gas entropy at $p^0 = 1$\,atm. It is listed in thermodynamic tables
\cite{CRC95} or can be easily obtained from first-principles calculations \cite{reuter02b}.

Choosing a pressure of $p_{\rm O_2} = 10^{-12}\,{\rm atm}$, characteristic for 
ultra-high vacuum (UHV) conditions, and using the above derived value for the 
binding energy, the $N_{\rm O}/N$ concentration is obtained as $6.2 \times 10^{-18}$ and $1.2 \times 10^{-14}$ for $T=300$\,K and $T=800$\,K, respectively. The number of octahedral interstitial sites in fcc Pd is $6.5 \times 10^{22}$\,sites/cm$^3$, which translates the above concentrations into $0.4 \times 10^6$ O-atoms/cm$^3$ for a temperature of 300\,K and to $0.8 \times 10^9$ O-atoms/cm$^3$ for 800\,K. Yet, most often, surface science experiments quantify uptakes in units of monolayers, which makes it desirable to convert the above numbers also into ML for direct comparison. Unfortunately, this requires to use a specified volume of the crystal. For convenience we take this to be $1$\,cm$^3$, which is of the same order of magnitude as a typical experimental sample. Then a coverage of 1\,ML corresponds to $1.5 \times 10^{15} {\rm atoms/cm^2}$ and the above stated concentrations translate into the equivalents of $10^{-10}$\,ML and $10^{-7}$\,ML at room temperature and $T=800$\,K, respectively. We conclude that the amount of bulk-dissolved oxygen in a macroscopic Pd crystal in equilibrium with its UHV surrounding is negligible.

Of course, during adsorption experiments, the crystal surface is exposed to much 
higher pressures than the UHV pressure considered above. On the other hand, it is unlikely that during the finite exposure time, the whole sample attains already the saturation concentration corresponding to full equilibrium with the surrounding higher gas pressure. If we therefore consider ambient pressure in our thermodynamic estimate, it will correspond to a conservative upper limit: The effective pressure during exposure will not exceed it, and the real uptake in the finite exposure time must be smaller than the equilibrium value. However, even when using $p_{\rm O_2} = 1$\,atm as the other extreme for the estimate in eq. (\ref{eq1}), we only obtain about $10^{-4}$\,ML ($10^{11}$ atoms/cm$^3$)
and $0.6$\,ML ($10^{15}$ atoms/cm$^3$) dissolved oxygen at room temperature and $T=800$\,K, respectively. We therefore always end up with uptake numbers that are significantly smaller than the few \cite{conrad77} to few hundred \cite{campbell76} ML of oxygen reported in experimental studies. Even considering the uncertainty due to the employed gradient-corrected DFT functional, the discrepancy between the experimentally reported oxygen uptakes and our calculated ones seems simply too large. We are therefore forced to conclude that the only place to store such large amounts of oxygen atoms in a Pd crystal is not in the regular interstitial sites considered here, but in bulk defects like grain boundaries and dislocations, or in oxidic structures presumably forming in or at the surface of the crystal. This conclusion is therewith equivalent to the one of a preceding study on the dissolution in Ru, where even smaller uptake numbers were computed \cite{reuter02a}. Considering that Pd is a much softer and larger lattice constant material with a correspondingly reduced lattice deformation cost upon O incorporation, the larger amount of oxygen dissolved in palladium is not astonishing. Unfortunately, corresponding studies have not yet been done for Rh and Ag, but one would expect the number of oxygen atom solved in Rh to be intermediate between the ones found for Ru and Pd, whereas the palladium value will be exceeded by Ag.

Returning finally to the initial motivation of studying bulk dissolution as  opposite limit to the immediate sub-surface incorporation studied in sections 
III A and B, we notice that the endothermic bulk binding energy of $E_b \approx -0.4$\,eV is comparable to the binding energies calculated for pure sub-surface oxygen adsorption, cf. Table I, i.e. for incorporation below the surface without adsorbed oxygen present on the surface (-/tetra-I). Both are therewith significantly less stable than adsorption in the mixed on-surface/sub-surface structures, leading us to conclude that O accommodation occurs preferentially in the immediate selvedge of the crystal. The ensuing strong local enrichment with oxygen entails then naturally the formation of ordered oxidic structures at the surface.

\section{Summary}

We presented a density-functional theory study addressing the initial oxygen penetration into the Pd(111) surface. The results show that sub-surface oxygen is initially always less stable than on-surface oxygen adsorption. However, with increasing coverage this preference vanishes, and oxygen incorporation becomes more favorable already above coverages of $\sim 1/2$\,ML. This threshold coverage is similar to the coverage for which bulk oxide formation becomes thermodynamically stable \cite{todorova02} and for which surface oxide formation was reported experimentally. We interpret this as reflecting a bottleneck function of the initial O penetration in the oxide formation process, with an almost immediate phase transition to oxidic structures after the first incorporation of oxygen below the surface. In the resulting structures with oxygen in on-surface and sub-surface interstitial sites, we always observe a clear preference for linear O-Pd-O arrangements, with occupation of fcc and tetrahedral sites. Similar {\em trilayer} structures had already been reported in preceding studies on the O incorporation into the basal surfaces of the other late $4d$ transition metals Ru, Rh and Ag. Analyzing the geometric and electronic structure of the fcc/tetra-I trilayer, we identify a good hybridization and screening within the trilayer, as well as an improved coupling to the underlying substrate as key factors towards the stabilization of this geometry. Comparing with bulk dissolved oxygen, our calculations indicate a strong enrichment of incorporated oxygen in the immediate selvedge of the crystal. Via the formation of trilayer structures, this suggests that the enrichment will then naturally entail the formation of oxidic structures at the surface.

\section{Acknowledgements}

The EU is acknowledged for financial support under contract no. NMP3-CT-2003-505670 (NANO$_2$).

\end{document}